\documentclass[usenatbib]{mn2e}
\usepackage{psfig}

\def\gs{\mathrel{\raise0.35ex\hbox{$\scriptstyle >$}\kern-0.6em
\lower0.40ex\hbox{{$\scriptstyle \sim$}}}}
\def\ls{\mathrel{\raise0.35ex\hbox{$\scriptstyle <$}\kern-0.6em
\lower0.40ex\hbox{{$\scriptstyle \sim$}}}}

\date{Accepted 2008 June 16.  Received 2008 June 13; in original form 2008 March 9}

\begin{document}

\title[KX-selected QSOs in the UKIDSS/UDS field]{A Pilot Survey for
KX QSOs in the UKIDSS Ultra Deep Survey Field}

\author[Smail et al.]
{Ian Smail,$^{\!1\,*}$
Rob Sharp,$^{\! 2}$
A.\,M.\ Swinbank,$^{\! 1}$
M. Akiyama,$^{\! 3}$
Y.\ Ueda$^{4}$ 
S.\ Foucaud,$^{\! 5}$
\and
O.\ Almaini,$^{\! 5}$ 
S.\ Croom$^{6}$
\vspace*{6pt} \\
$^1$Institute for Computational Cosmology, Department of
Physics, Durham University, South Road, Durham, DH1 3LE, UK \\
$^2$Anglo-Australian Observatory, Epping, NSW 1710, Australia\\
$^3$Subaru Telescope, National Astronomical Observatory of Japan, 650 North A'ohoku Place, Hilo, HI 96720, USA\\
$^4$Department of Astronomy, Kyoto University, Kyoto 606-8502, Japan\\
$^5$School of Physics and Astronomy, University of Nottingham,
University Park, Nottingham NG7 2RD\\
$^6$School of Physics, University of Sydney, NSW 2006, Australia\\
$^*$Email: ian.smail@durham.ac.uk \\
}

\maketitle

\begin{abstract}
We have undertaken a  pilot survey for faint QSOs in the UKIDSS Ultra
Deep Survey Field using the KX selection technique. These observations
exploit the very deep near-infrared and optical imaging of this field
from UKIRT and Subaru to select candidate QSOs based on their $VJK$
colours and morphologies.  We determined redshifts for 426 candidates
using the AAOmega spectrograph on the AAT in service time.  We identify
17 QSOs ($M_B\ls -23$) in this pilot survey at $z=1.57$--3.29.  We
combine our sample with an X-ray selected sample of QSOs in the same
field (a large fraction of which also comply with our
KX selection) to constrain the surface density of QSOs with $K\leq 20$,
deriving limits on the likely surface density of 85--150\,deg$^{-2}$.
We use the good image quality available from our near-infrared imaging
to detect a spatially extended component of the QSO light which
probably represents the host galaxies. We also use our sample to
investigate routes to improve the selection of KX QSOs at faint limits 
in the face of the significant contamination
by compact, foreground galaxies. The brightest examples from our
combined QSO sample will be used in conjunction with a large VLT VIMOS
spectroscopic survey of high redshift galaxies in this region to study
the structures inhabited by gas, galaxies and growing super-massive
black holes at high redshifts in the UKIDSS UDS.
\end{abstract}


\section{Introduction}

Simulations of the distribution of baryons in the Universe predict that
much material is spread out in a ``cosmic web'' between the galaxies,
groups and larger structures visible in conventional surveys.  The main
components of this material are the Ly$\alpha$ forest, which traces
highly ionized hydrogen of low neutral column density and low chemical
enrichment distributed through the lower density regions, and metal
absorption lines which trace massive galaxy halos (and hence higher
density regions) via their metal-enriched gas. To probe this material
and investigate its 3-dimensional distribution we must exploit QSOs as
bright background sources to trace the web through its absorption.

An optimal survey of the cosmic web would use a grid of distant bright
QSOs as probes of the intervening matter, with the redshift and
equivalent width of these absorbers providing information on the
relative distribution and clustering of gas around galaxies and larger
structures, as well as providing insights into the chemical enrichment
and heating of this material (e.g.\ Morris \& Januzzi 2006).  Studies
comparing the results from high-redshift galaxy surveys around
individual bright QSOs have provided unique insights into the effect of
star formation on the gas surrounding young galaxies (e.g.\ Adelberger
et al.\ 2005).  However, these studies are observationally expensive --
it would be much more efficient if a large number of galaxies and QSOs
could be compared within a single field as the number of
absorber--galaxy pairs scales as N$^2$.  The difficulty is that the
surface density of bright QSOs is low and hence faint QSOs are required
to set up a dense grid of probes needed to sample the full range of
cosmic structures in a single field (Prescott et al.\ 2006).  Nevertheless,
with enough QSOs in a single field the long integration time required
to undertake absorption-line analysis on relatively faint background
sources (AGN or galaxies) become worthwhile.

We are undertaking a highly-sampled survey of the cosmic web in a
uniquely well-studied field: the UKIDSS (Lawrence et al.\ 2007)
Ultra Deep Survey (UDS) field. The UDS is the deepest, panoramic
near-infrared observations so far undertaken with the goal of reaching
$K\sim 23$ and $J\sim 25$ over $\sim 0.8$ sq.\ degrees (Foucaud et al.\
2007).  Equally deep $BV\! RIz$ imaging is available across the whole
field from Subaru/XMM Deep Survey (SXDS, Furusawa et al.\ 2008) and this
region has also been extensively studied at X-ray wavelengths as
part of the same project (Ueda et al.\ 2008).
These data are being analysed to study the build-up of massive galaxies
out to $z\sim3$ and its variation with environment (Foucaud et al.\
2007; Lane et al.\ 2007; Cirasulo et al.\ 2007).  This field is also
the target of a VLT Large Programme (PI: O.\ Almaini) to use the VIMOS
spectrograph to obtain redshifts for $\sim 5$,000 high-redshift
$K$-selected galaxies.  In this paper we present a parallel pilot
survey to expand the size of the QSO sample in this field to allow us
to relate these luminous galaxies and structures to their surrounding
enriched and unenriched gas from a highly-sampled grid of sightlines.
Combining these two unique datasets will yield powerful insights into
the influence of galaxies and AGN (at the peak era in their activity)
on low-density gas in their environments, material which is required
for the long-term fueling of the star formation in galaxies.

To construct a highly-sampled grid of QSOs in this field we have used
$VJK$-band imaging from UKIDSS/UDS to apply the KX method (Warren et al.\
2000; Croom et al.\ 2001; Sharp et al.\ 2002) to identify a sample of
$\sim$\,800 $V<23$ potential AGN candidates in this region.  The good seeing
available in the UDS imaging, 0.8$''$ FWHM across the whole field,
yields relatively clean catalogs of point sources -- which are a mix of
AGN, stars and compact galaxies.  The KX method relies on the power-law
shape of AGN continuua across the $VJK$ bands, compared to the $H$-band
``bump'' arising from the opacity minimum in stellar atmospheres, to
distinguish between those point sources which are AGN and those which
are stars.  This color selection has the additional advantage that it
provides an unbiased selection of QSOs independent of their dust
reddening out to high redshifts (Warren et al.\ 2000), compared to the
classical ``UVX'' selected samples.

We use a cosmology with $H_{0}=70$\,km\,s$^{-1}$, $\Omega_{M}=0.3$ and
$\Omega_{\Lambda}=0.7$ in which 1$''$ corresponds to 8.2\,kpc at
$z=2.5$ and 8.5\,kpc at $z=1.5$.  All quoted magnitudes are  Vega
and as the galactic reddening in our field is low, $E(B-V)=0.02$,
we have not applied any extinction corrections to the colours or magnitudes
in this paper.

\section{Observations and Reduction}

%
%
\begin{figure}
\centerline{\psfig{file=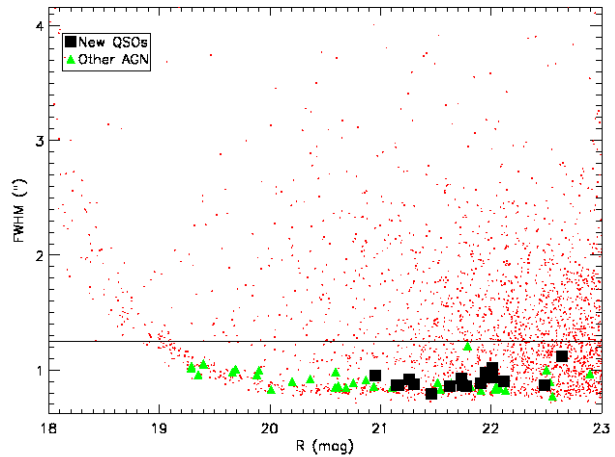,angle=90,width=4.0in}}
\caption{A plot of the FWHM in the $R$-band versus $R$-band magnitude
for a random 1-in-3 subset of sources in the UKIDSS UDS. The stellar
sequence is visible at $\sim 0.8''$ (image saturation causes the turn
up in the sequence at bright magnitudes).  To ensure that we are not
biased against QSOs with bright host galaxies or very close companions
-- we broaden our selection to FWHM\,$\leq 1.25''$.  We also plot the
new QSOs identified from our survey, which show a modest range in $R$-band
FWHM, as well as the existing X-ray selected AGN which comply with our
selection and which were used to tune the criteria.
}
\end{figure}

\subsection{Photometric Selection}

Our sample selection starts from the matched near-infrared/optical
catalogue of sources in the UKIDSS UDS field.  The near-infrared data
come from the UKIDSS Data Release 1 (Warren et al.\ 2007) and the
optical photometry are from an early version of the Subaru catalogues
(O.\ Almaini, priv.\ comm.)  published by Furusawa et al.\ (2008). 
The combined catalogue comprises a seeing-matched 
$K$-band selected sample which is then matched to the seeing-matched
optical Subaru catalogues (see Furusawa et al.\ 2008) through the $i$-band.  The
methods used to create this combined
optical--near-infrared photometric catalogue are described more fully
in Foucaud et al.\ (2007) and the DR1 catalogue itself is
described Almaini et al.\ (in prep). For our analysis we
select only those objects lying in areas with good photometry, without
bright neighbours and which are unsaturated/etc, leaving us with over
36,000 objects with $BV\! RIzJK$ photometry across a 0.572 degree$^2$
field.  The 3-$\sigma$ photometric limit of the catalogue in our
key bands are $V=27.8$, $R=27.0$, $J=23.4$ and $K=22.2$.

In the following we also employ the existing spectroscopic survey of
X-ray sources in the SXDF/UDS from Akiyama et al.\ (2008).  This provides
a useful sample of known AGN to assess our selection
criteria. The Akiyama et al.\ survey targets the 952 X-ray sources detected either in the 0.5--2 or
2--10\,keV bands from the area of the {\it XMM-Newton} survey of the SXDF (Ueda et al.\ 2008)
which is covered by the deep optical imaging
in Furusawa et al.\ (2008). Of these 952 X-ray sources, 648 have optical
counterparts brighter than $R_{AB}=24$ in Furusawa et al.\ (2008). Intensive optical spectroscopic observations of these optical counterparts have been performed
with Subaru/FOCAS and VLT/VIMOS and to date $\sim 60$\% of the 648 X-ray sources
have been spectroscopically identified.  The majority of these are moderate redshift
AGN, with a tail of QSOs extending out to beyond $z\sim 5$.
As we discuss below, in addition to using this catalogue as a training set for our
sample selection, we have culled from our catalogue any source with spectroscopy
in Akiyama et al., although we have include  all the QSOs with 
$M_B\leq -23.0$ and brighter than $K_{tot}=20.0$ from their survey
in our final estimate of the total surface density of QSOs in the $K$-band.

We first identify the locus of stars in the FWHM--magnitude plane.  We
use the Subaru $R$-band FWHM for this purpose for two reasons.
Firstly, we do not want to bias our selection against QSOs with
detectable host galaxies, as these hosts are likely to be redder than
the QSOs this argues for using the bluer optical bands in the
morphological selection, rather than our near-infrared data.  Secondly,
the image quality in the $R$-band provides the best combination of
image quality and depth from our available data.  We plot $R$-band FWHM
versus $R$-band magnitude in Fig.~1 and show our FWHM\,$\leq1.25''$
selection limit.  This limit was chosen by comparison to the existing
spectroscopic sample in this field, with the aim of yielding a sample
well-matched to the grasp of the AAOmega spectrograph with modest
contamination.  Setting the FWHM limit to $\leq 1.0''$ yielded a sample
of 218 sources complying with our KX-selection (see below).  Of these,
43 matched sources in the existing spectroscopically identified
catalogue (Akiyama et al.\ 2008) of which 41
were classified as broad emission line sources (although not all of 
these have $M_B\leq -23$, our definition of a QSO), with just 2 were narrow
emission line sources.  This selection of AGN thus appeared very pure,
but is not necessarily complete and moreover would not fully populate
the AAOmega fibres.  Extending the FWHM cut to $\leq 1.25''$ increased
the sample of sources with the appropriate colours to 758 (well matched
to AAOmega), yielding 89 matches in the existing spectroscopic catalog,
from which 47 were broad emission line sources, 21 are narrow emission line sources and the
remainder were classified as absorption line spectra.  Thus a cut on
$R$-band FWHM at $\leq 1.25''$ yielded a sufficiently large sample of
KX-selected targets to fill all the AAOmega fibres with a moderate
level of contamination (estimated from the existing spectroscopic
sample).  This relatively relaxed constraint on image size also ensures
that we do not reject QSOs in which the host galaxy is detectable.

We then use the 2$''$-diameter aperture photometry in the $V\! JK$
bands to isolate the stellar sequence in colour space (Fig.~2).  We
limit our sample to $K_{tot}\leq 22.0$ to ensure completeness in the
$K$-band catalogue and then required targets complied with: $V_{ap}\leq
23.0$; $(J-K)_{ap}\geq 1.10$; $(J-K)_{ap}\geq 0.25*(V-J)_{ap}+0.40$.
This selection is a variation on that used by Croom et al.\ (2001),
where the changes reflect small differences in the magnitude systems.
We note that as our photometric catalogue is seeing-matched, we do not
apply any aperture corrections to any of the colours in this paper.  We
have also ignored the effect of variability on our measured colours and
magnitudes, and caution that this may be responsible for throwing QSOs
out of our selection.  Finally, we also stress that given the typical
colours of QSOs, $(V-K)\sim 2.5$--3.5, the relatively bright $V$-band
magnitude limit required by our spectroscopic follow-up means that the
QSO sample is effectively limited at $K\sim 20$ so all have very
good $K$-band detections, $>20$-$\sigma$.

As a final step before our spectroscopic observations, we remove from
our sample the 89 objects which are in the existing redshift catalog of
Akiyama et al.\ (2008), including their X-ray QSOs  which comply with our
KX-selection and have $M_B\leq -23.0$, which we discuss further in
\S3.2.  After removing these sources we have a final input catalogue
comprising 670 targets.  In our subsequent spectroscopy we gave
priority to sources with $V<22.0$ above those with $V<22.5$ and
$V<23.0$.  We model the effects of this variable completeness on our
results.

%
%
\begin{figure}
\centerline{\psfig{file=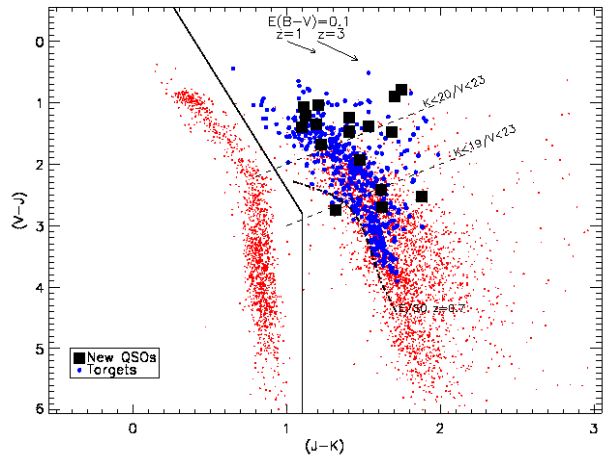,angle=90,width=4.0in}}
\caption{$(V-J)$ versus $(J-K)$ for objects from the UKIDSS UDS.  We
plot as the background the distribution of colours for 
the $K_{tot}\leq 20.0$ sources in our field.  The stellar
sequence is clearly visible and can be fairly well separated from the AGN
(and compact galaxies) using the colour selections marked.  We identify
the subset of sources selected on the basis of their morphologies and
colours to be targeted for AAOmega spectroscopy and the 17 new QSOs which
are identified by this spectroscopy. As can be seen these QSOs are well
distributed across the colour plane.  We also indicate the potential
influence of dust reddening (either intrinsic or from foreground
absorption systems) on the colours of QSOs at $z\sim 1$--3.  We also
mark on the selection boundaries (dashed lines) for sources with $V<23$ and either $K<20$
or $K<19$, and the expected variation in apparent colour of
a galaxy with a non-evolving E/S0 SED from $z=0$ (top) to $z=0.7$ (dot-dashed line).
}
\end{figure}

%
%
\begin{figure*}
\centerline{\psfig{file=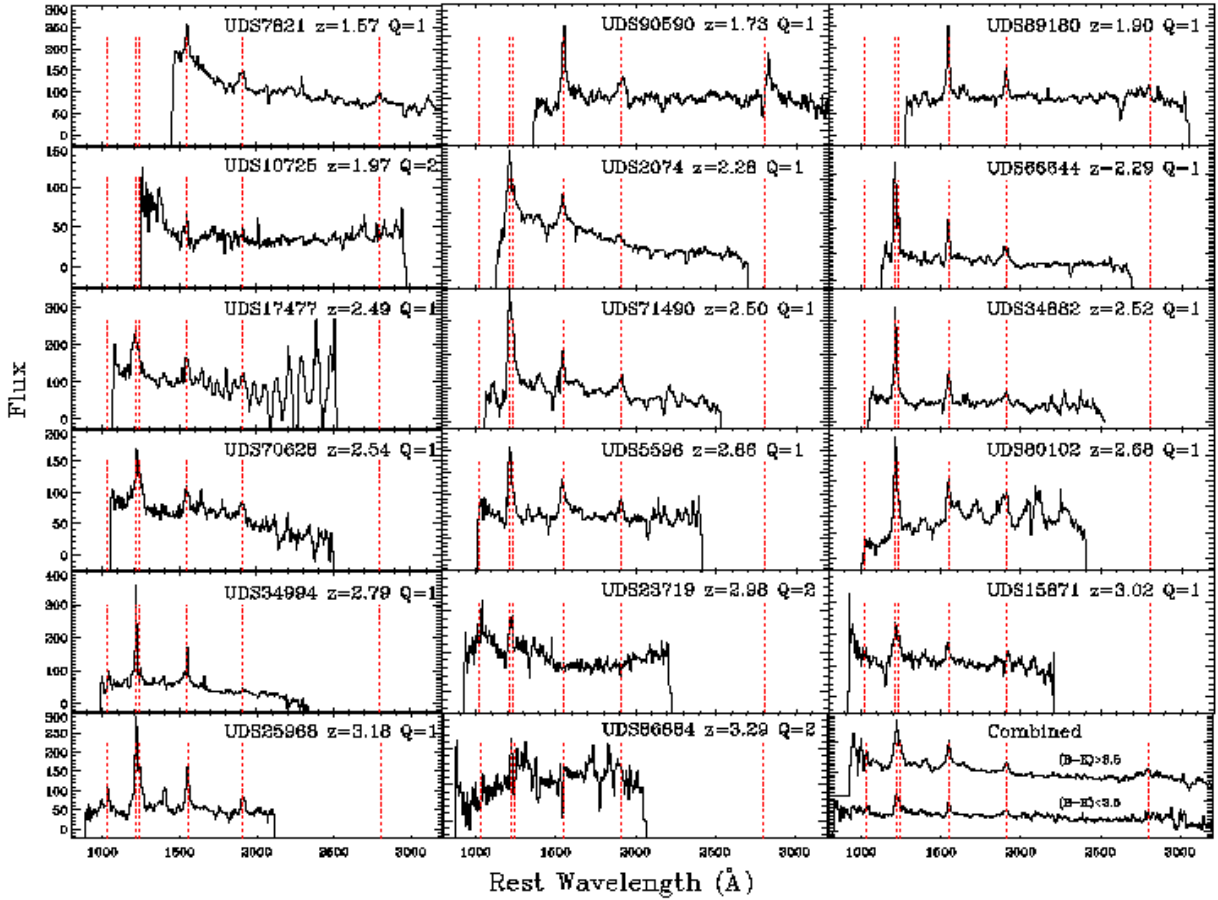,angle=90,width=7.0in}}
\bigskip
\caption{Restframe spectra of the 17 new KX-selected QSOs from the AAOmega
observations of the UKIDSS/UDS.  These are order in increasing redshift
(from the upper-left) and have been smoothed with a 7\AA-FWHM Gaussian
for display purposes. 
In the final panel we show two combined spectra
for the redder and bluer QSOs in the sample, divided at $(B-K)=3.5$.
We mark on each spectrum the expected wavelengths of common emission
features: Ly$\beta$\,$\lambda$1026, Ly$\alpha$\,$\lambda$1216, N{\sc
v}\,$\lambda$1243, C{\sc iv}\,$\lambda$1549, C{\sc iii}]\,$\lambda$1908
and Mg{\sc ii}\,$\lambda$2800. The red end of the spectrum of UDS17477
shows interference fringing at the fiber/prism interface on the 2dF
fibre positioner.  We stress that the spectra have not been accurately
flux calibrated (\S2.2) and so the continuua shape should be treated
with caution. 
}
\end{figure*}

\subsection{AAOmega Observations}

The spectroscopic observations were undertaken as part of the service
programme on the Anglo-Australian Telescope (AAT).  The total
integration time was 6\,hrs spread equally over three nights (2006
Aug.\ 31, Sept.\ 28 and 2007 July 15).  The conditions for the first
two runs were good, with seeing $<1.5''$ FWHM and good transparency,
although the third run suffered from poor seeing, $3.0''$ FWHM, and
yielded few additional redshifts.

Our observations used the AAOmega spectrograph and the multi-object
fibre feed from the 2dF fibre positioner system (Saunders et al.\ 2004;
Smith et al.\ 2004; Sharp et al.\ 2006).  The dual-beam AAOmega
spectrograph was used in it's default low resolution configuration, the
580V and 385R VPH gratings delivering a 3.4 pixel resolution element
and R$\sim$1300 over the wavelength range $\lambda= 3700$--8600\AA\ (the
lower limit essentially set by the transmission limit of the 38 meter
fibre optic feed, the upper limit constrained by the available CCD
coverage).  The 5700\AA\ dichroic mirror was used to separate the twin
beams, the change over occurring in the region $5700\pm200$\AA.

The observations where broken up into 2 hour blocks to minimize losses
from atmospheric effects\footnote{For details on these considerations
see http://www.aao.gov.au/AAO/2df/aaomega/aaomega\_CVD.html}.  Each
observation block consists of a quartz-halogen flat-field exposure
(used both to trace fibre footprint on the CCD and also for the 
relative response of the fibres), a composite arc lamp frame (utilizing
CuAr, FeAr hollow cathode arc-lamps and neutral density filtered Helium
and Neon lamps) used for primary wavelength calibration, and four
science exposures (each of 1800\,s).  Twenty five of the $\sim$370
fibres available on the 2dF positioner at the time of observation were
allocated to sky positions for sky subtraction.  These positions were
hand-picked from the Subaru imaging.  The contemporaneous sky
observations are reduced as part of the science fibre reduction, and
then combined to make a high signal-to-noise sky spectrum free from
cosmic rays and CCD defects for each science exposure.

As is usual for AAOmega observations, the data where processed using
the {\sc 2dfdr} data reduction package\footnote{Available from the AAO
web site at
http://www.aao.gov.au/AAO/2df/aaomega/aaomega\_software.html}.  Data
from the blue and red arms of the spectrograph are processed
independently.  In addition to the processing steps common to all fibre
spectrograph systems (overscan correction, fibre \emph{tramline map}
generation, spectral extraction, fibre relative response and wavelength
calibration and sky subtraction) {\sc 2dfdr} implements a Laplace
filtering cosmic ray rejection to identify cosmic rays before spectral
extraction (van Dokkum 2001) and an iterative sky subtraction involving
minimizing the sky subtraction residual by iteratively scaling for:
small relative wavelength shifts between the composite sky spectrum and
each science spectrum; slight degradation of the science spectrum
resolution to match the composite sky; and the intensity of the sky
spectrum.

The multiple science spectra for each object are then combined using a
weighting derived from the average flux recorded in a subset of the
brighter objects in each field (to account for seeing and transparency
variations during observation).  The blue and red arm data were then
spliced together to give a continuous spectrum using an archival
transfer function to correct for broad band sensitivity. No detailed
flux calibration is attempted.  Data from all previous nights were then
stacked to build up signal on the fainter targets for which redshift
had not been determined.  Between each night, all low-redshift ($z\ll
1$) objects for which redshifts were reliably measured were replaced in
subsequent observations with alternate targets.  This maximized the
yield for the final science exposure.  In total 626 sources from our
candidate catalogue of 670  were observed.

\begin{table*}
\begin{center}
{\footnotesize
{\centerline{\sc Table 1.}}
{\centerline{\sc Properties of new KX-selected QSOs}}
\begin{tabular}{lcccccccccc}
\hline
\noalign{\smallskip}
ID & R.A.\ & Dec.\ & $K_{tot}$ & $(B-V)_{ap}$ & $(V-J)_{ap}$ & $(J-K)_{ap}$ & $z$ &
M$_B$ & $f_X^a$ & $L_X^b$ \\
   & \multispan2{ ~(J2000) }  &&&&&&&   (0.5--4.5\,keV) &  (2--10\,keV)\\
\hline
UDS2074  & 02\,17\,03.86 & $-05$\,31\,40.6 & 18.64& $-0.00\pm 0.01$~ & $1.47\pm 0.01$ & $1.40\pm 0.14$ & 2.286& $-24.2$ & $5.9\pm  0.9$ &1.9\\
UDS5596  & 02\,19\,10.00 & $-05$\,29\,30.8 & 18.62& $0.43\pm 0.01$ & $1.22\pm 0.01$ & $1.13\pm 0.16$ & 2.665& $-24.2$ & $2.6\pm  0.8$&1.1\\
UDS7821  & 02\,18\,45.94 & $-05$\,28\,06.5 & 18.47& $0.01\pm 0.01$ & $1.39\pm 0.01$ & $1.53\pm 0.14$ & 1.570& $-23.2$ & $8.4\pm  0.9$&1.0 \\
UDS10725$^*$ & 02\,18\,50.17 & $-05$\,26\,22.6 & 18.63& $0.60\pm 0.02$ & $2.41\pm 0.02$ & $1.61\pm 0.15$ & 1.979& $-23.3$ & $<1.5$&$<0.8$        \\
UDS15871 & 02\,18\,47.62 & $-05$\,23\,19.9 & 19.95& $0.75\pm 0.01$ & $1.08\pm 0.03$ & $1.11\pm 0.26$ & 3.020& $-23.1$ & $2.6\pm  0.5$&1.4\\
UDS17477 & 02\,17\,30.58 & $-05$\,22\,22.9 & 19.08& $0.41\pm 0.01$ & $0.90\pm 0.02$ & $1.70\pm 0.18$ & 2.498& $-23.3$ & $5.7\pm  0.6$&2.0\\
UDS23719$^*$ & 02\,18\,27.47 & $-05$\,18\,37.2 & 18.19& $0.83\pm 0.02$ & $2.52\pm 0.02$ & $1.88\pm 0.13$ & 2.988& $-24.4$ & $<1.3$&$<1.1$\\
UDS25968 & 02\,16\,20.71 & $-05$\,17\,18.7 & 18.85& $1.24\pm 0.01$ & $1.36\pm 0.01$ & $1.19\pm 0.14$ & 3.190& $-23.7$ & $7.7\pm  1.2$&7.6\\
UDS34882 & 02\,19\,24.29 & $-05$\,11\,49.6 & 18.66& $0.42\pm 0.01$ & $1.19\pm 0.02$ & $1.83\pm 0.10$ & 2.522& $-23.4$ & $6.3\pm  1.3$&6.4\\
UDS34994 & 02\,16\,55.23 & $-05$\,11\,46.7 & 19.50& $0.29\pm 0.01$ & $1.04\pm 0.02$ & $1.21\pm 0.19$ & 2.793& $-23.4$ & $3.4\pm  0.7$&1.5\\
UDS66644 & 02\,18\,08.62 & $-04$\,53\,54.6 & 18.35& $0.55\pm 0.01$ & $1.48\pm 0.01$ & $1.68\pm 0.12$ & 2.296& $-23.8$ & $<0.9$&$<0.6$\\
UDS70628 & 02\,18\,40.26 & $-04$\,51\,46.6 & 19.39& $0.34\pm 0.01$ & $0.78\pm 0.03$ & $1.74\pm 0.18$ & 2.550& $-23.2$ & $6.2\pm  1.0$&2.3\\
UDS71490 & 02\,17\,13.13 & $-04$\,51\,15.9 & 18.43& $0.40\pm 0.01$ & $1.24\pm 0.02$ & $1.41\pm 0.15$ & 2.504& $-24.2$ & $<2.1$&$<1.5$\\
UDS80102 & 02\,17\,14.04 & $-04$\,46\,12.6 & 18.70& $0.66\pm 0.01$ & $1.69\pm 0.01$ & $1.22\pm 0.15$ & 2.690& $-24.0$ & $4.2\pm  0.6$&1.7\\
UDS86884$^*$ & 02\,17\,35.95 & $-04$\,42\,33.6 & 18.77& $1.17\pm 0.02$ & $2.75\pm 0.02$ & $1.32\pm 0.15$ & 3.292& $-24.1$ & $1.7\pm 0.5$&1.4\\
UDS89180 & 02\,17\,32.87 & $-04$\,41\,17.6 & 19.15& $0.34\pm 0.01$ & $1.40\pm 0.02$ & $1.10\pm 0.18$ & 1.907& $-23.3$ & $3.8\pm  0.5$&0.7\\
UDS90590 & 02\,16\,47.28 & $-04$\,40\,30.0 & 18.76& $0.28\pm 0.01$ & $0.95\pm 0.02$ & $1.46\pm 0.10$ & 1.728& $-23.1$ & $7.7\pm  0.8$&1.5\\
\hline
\hline
\end{tabular}
\caption{Typical errors are $\pm 0.5''$ on positions, $\pm 0.05$ on
  total $K$-band magnitudes, $\pm 0.1$--0.3 on $M_B$ and
  $\pm 0.001$ on $z$.  All redshifts have Q=1 except for those marked
  $^*$ which have Q=2.  a) units are 10$^{-15}$
   erg\,s$^{-1}$\,cm$^{-2}$. b) units are 10$^{44}$\,erg\,s$^{-1}$.
}

}
\end{center}
\medskip
\end{table*}

\section{Analysis, Results and Discussion}

\subsection{Spectroscopic Measurements}

Redshift determination was performed using the {\sc autoz} package,
which is optimized for measuring QSO redshifts (Croom et al.\ 2001)
and the {\sc runz} package developed for the 2dFGRS (Colless et al.\
2001) and updated for the 2SLAQ project (Cannon et al.\ 2006).  All
redshifts were then visually checked.  From the 626 objects observed,
our spectra yield a total of 426 sources with redshifts of quality (Q)
Q=1, unambiguous, or Q=2, probable.  Of these 426, 11 are galactic
stars:  4 DA white dwarfs, 2 DB white dwarfs and  five main sequence stars 
(one each of A, B, F, G and M). Thus the colour selection from Fig.~2 has
significantly reduced, but not totally eliminated, the stellar contamination.

We derive the absolute restframe $B$-band magnitudes for these extragalactic sources
using {\sc kcorrect} (Blanton \& Roweis 2007) by interpolating based on
a fit to the optical and near-infrared photometry spanning the
restframe $B$-band.  These absolute magnitudes are based on the
aperture
photometry and so we correct them to total magnitudes using the mean
aperture correction for the sample, $\delta = -0.4\pm 0.1$, derived
in the $J$-band photometry.
We then apply a $M_B\leq -23.0$ cut to identify the
QSOs from our spectroscopic sample. This results in 17 QSOs
with $M_B\sim -23.3$ from our AAOmega observations in the UDS which
comply with our KX and various image selection criteria.  Their
redshifts span $z=1.57$--3.29 and we show the spectra for these in
Fig.~3 and list their positions and photometric properties in Table~1.
The X-ray fluxes in the table come from Ueda et al.\ (2008) and are in
the observed 0.5--4.5\,keV band (assuming a power-law photon index of
1.8 with no absorption).

As expected, the spectra of the QSOs in our sample were all best
matched with the QSO (broad emission line) template spectrum employed by {\sc runz}
and they typically display broad emission lines, including
Ly$\alpha$\,$\lambda$1216, N{\sc v}\,$\lambda$1243, C{\sc
iv}\,$\lambda$1549, C{\sc iii}]\,$\lambda$1908 and Mg{\sc
ii}\,$\lambda$2800.  In the lower-right panel of Fig.~3 we show the
composite spectra formed by combining the individual spectra for the
redder and bluer QSOs (divided at $(B-K)=3.5$) to demonstrate their
common features.  We note that the redder QSOs show proportionally
higher equivalent width emission lines than the blue QSOs.  Given the
small sample available, this could result from: 1) a selection effect
as the bluer QSOs tend to be lower redshift; 2) a selection effect
arising from the need for stronger features to be present in the redder
QSOs, which are typically fainter in the wavelength range of our
AAOmega observations; or 3) absorption and reddening of the underlying
continuum in the redder QSOs.

The mean redshift for the new KX QSOs in our survey is $z=2.50\pm 0.50$.
This compares to a redshift range of $z=1.08$--4.55 and a mean of
$z=2.05\pm 0.71$ for the 23 X-ray selected sources from Akiyama et al.\
(2008) which comply with our KX selection and $M_B$ absolute magnitude
limit, but which were removed from our initial sample selection.  As expected the
redshift distributions for these two samples are statistically
indistinguishable and the combined sample has a mean redshift of
$z=2.30\pm 0.81$.

%
%
\begin{figure}
\centerline{\psfig{file=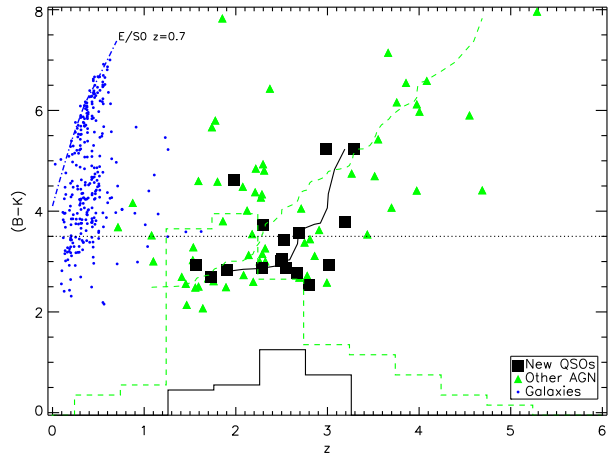,angle=90,width=3.5in}}
\caption{The variation in $(B-K)$ colour as a function of $z$ for the
QSOs in the UDS, as well as the galactic contaminants.  We  plot the
trend in the median colour for the AAOmega KX-sample (solid) and for the
X-ray selected QSOs in the field (dashed).  For comparison to the
galaxy colours we plot the track of a non-evolving E/S0 galaxy
(dot-dashed), which defines the red-envelope of the low-redshift
contamination.  The horizontal dotted line shows the dividing line for
``red'' QSOs, with $(B-K)\geq 3.5$: $\sim 35$\% of the new KX-selected
sample are redder than this limit. The histograms at the bottom of the figure show
the redshift distribution for the KX-selected QSOs from our AAOmega
observations and the existing X-ray selected QSO sample in the UKIDSS UDS field.
}
\end{figure}

\subsection{QSO Surface Density}

We can use our survey to place a lower limit on the number density of
QSOs brighter than our effective magnitude limit of $K_{tot}\leq 20$.
Correcting for the weighting in our spectroscopic sample selection
suggests our KX selection would have yielded more than 17 QSOs if we
had observed all of the candidates (this does not include QSOs where
the spectra failed to yield an ID).  This correction implies 18.2 QSOs in our
0.572 degree$^{2}$ field.  

To determine the total number density of QSOs in the field, we have to
add in the 30 $M_B\leq
-23.0$ QSOs brighter than  $K_{tot}=20.0$ which were removed from our target list as
they had existing spectroscopy from Akiyama et al.\ (2008).
These include the 23 QSOs which comply with
our KX and morphological selection, and a further seven which
are not identified by the KX
colour and morphology selections.  Four of these seven are more extended
(in $R$, and $K$) than our FWHM cut, one falls outside the KX color
region and two fall in regions with poor photometric coverage.
Combining these two samples yields a surface density of $\geq 85$ QSOs per degree$^{-2}$
brighter than $K_{tot}=20.0$.

The main uncertainty in the estimate above is the incompleteness in our
spectroscopic identifications, which is significant for $V>22$ sources.
Correcting for this incompleteness is highly uncertain as it requires
an assumption about the spectral properties of any undetected QSOs.
Assuming that all of these systems exhibit strong emission lines would
suggest little incompleteness in our identification of QSOs, leading to
an actual surface density close to the limit quoted above.  However,
such an assumption is at odds with the spectral properties of even the
detected QSOs (Fig.~3).  An alternative approach is to assume that the
incompleteness of the QSO sample is similar to the overall
incompleteness of our identifications for the full spectroscopic sample
at the same $V$-band magnitudes (i.e.\ that the incompleteness simply reflects
continuum signal-to-noise in the spectra).  Adopting this assumption we
estimate a surface density of QSOs with $K_{tot}\leq 20.0$ in our
survey of 56 degree$^{-2}$ and correcting for our sampling and adding
in the known X-ray selected QSOs, raises this to $\sim 150$
degree$^{-2}$.

Our limit on the QSO surface density are broadly consistent with the
published limit of 325$^{+316}_{-177}$\, deg$^{-2}$ at $K\ls 19.5$ by
Croom et al.\ (2001), based on the detection of 3 QSOs in a small
48\,arcmin$^2$ field.  As expected, our observed QSO surface density is
significantly higher than that estimated at $K\sim 15$ by Glikman et
al.\ (2004, 2007), of just $\sim 0.1$ deg$^{-2}$.  A more useful
comparison is to the recently completed $K$-band QSO survey by Maddox
et al.\ (2008).  They surveyed 12.8 degree$^2$ to $K=17$ and determine
a surface density of $15.3\pm 1.1$ deg$^{-2}$, indicating that QSO
counts increase roughly linearly with $K$-band flux fainter than
$K=17$.

In terms of theoretical predictions, Maddox \& Hewett (2006) provide
estimates for the number counts of QSOs in the $K$-band based on a
standard model of the evolution of the QSO Luminosity Function, but
with the addition of a model for the influence of host galaxy light on
the apparent magnitudes of the QSOs.  Their pure-QSO model (no
contribution from the host galaxy) predicts $\sim 100$ degree$^{-2}$
QSOs with $M_B\leq -23.0$ QSOs and $K\leq 20.0$, while the inclusion of
light from the host galaxy in their models increases this to $\sim 110$
degree$^{-2}$.  We therefore conclude that our observational limits are
consistent with the predicted number counts of QSOs in the
$K$-band from Maddox \& Hewett (2006) and depending upon the degree of
incompleteness in our survey may support a modest contribution from the
host galaxies in the $K$-band light from our QSO sample.

\subsection{QSO Properties}

%
%
\begin{figure}
\centerline{\psfig{file=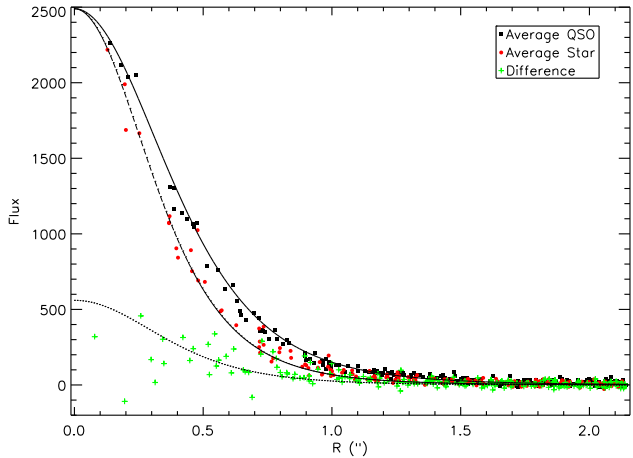,angle=90,width=3.5in}}
\caption{The radial $K$-band light profiles for the average QSO in our
KX-selected sample and a spatially- and magnitude-matched sample of
colour-selected stars.  The measured FWHM from Moffat fits to the
average QSO and star are $0.80\pm 0.05''$ and $0.67\pm 0.05''$
respectively, showing that the QSOs are on average more extended than
nearby stars.  Subtracting a scaled star from the composite QSO suggest
the residual light comprises 25--40\% of the light of the QSO,
equivalent to $K_{tot}\sim 20.3$, and has an extent of several kpc.
The profiles of the average QSOs which are redder and bluer than
$(B-K)=3.5$ indicates that the bluer QSOs are more compact, $0.75\pm
0.05''$, although this is not statistically significant.
}
\end{figure}

We find a range of $(B-K)= 2.5$--5.5 for our QSO sample (Fig.~4), less
than found for lower-$z$ QSOs (Glikman et al.\ 2004).  This moderate
range in colour likely reflects the limit imposed by our spectroscopic
incompleteness which becomes substantial beyond $V=22$ (equivalent to
$K\sim 18$), with only the bluest QSOs having adequate $V$-band signal-to-noise for
us to measure a redshift at $K\sim 20$.  Nevertheless, we have 6/17
QSOs ($\sim35$\%) in our sample redder than the $(B-K)=3.5$, frequently
used to identify red QSOs (e.g.\ Jurek et al.\ 2008).  Although we note
that we have no examples of very red QSOs with $(V-J)\geq 3$ in our
AAOmega sample (Croom et al.\ 2001), again likely due to the
incompleteness for fainter $V$-band sources in our spectroscopy,
but such red QSOs do exist in the X-ray sample in this field.

The new KX-selected sample of QSOs in Fig.~4 displays a trend to redder $(B-K)$
colours for higher redshift QSOs. This most likely reflects increasing
absorption by the Ly$\alpha$ forest effecting the $B$-band emission of
more distant QSOs (the X-ray sample exhibits a similar trend, Fig.~4).

We can exploit the high-quality multiwavelength data for the UDS region
to investigate the properties of the QSOs we have discovered.  Of our
17 new QSOs, 13 fall within 5$''$ of X-ray sources in the {\it
XMM-Newton} catalogue of this field (Ueda et al.\ 2008) and we list
their X-ray fluxes in Table~1.  Indeed, we note that 31 sources from
our spectroscopic sample of 426 objects are matched to the 952 sources
in the X-ray catalog within 5$''$, and that of the 13 of these X-ray
sources which lie at $z>1.5$, 12 appear in our KX QSO catalogue.  The
thirteenth was removed as the spectra were of too low quality.  For the
13 X-ray detected QSOs we derive colours in the observed 0.5--4.5\,keV
band for these sources. For 12/13 their colours are consistent with no
or little absorption, indicating that these are all type-1 AGNs.
However, the X-ray color of UDS34882 appears hard in the 0.5--4.5\,keV
band, and suggests an absorbing column of N$_H$ of $2 \times 10^{23}$\,cm$^{-2}$
(assuming a photon index of 1.8). This source therefore appears to be a
candidate of an ``X-ray absorbed'' QSO.

Comparing the redshift distributions of the X-ray undetected and
detected subsamples in Table~1, we see that the X-ray undetected and detected QSOs
have similar redshifts distributions ($z=2.44\pm 0.18$ versus
$z=2.52\pm 0.15$, where the errors are bootstrap estimates).  Looking
at the X-ray luminosities and limits of the detected and undetected
QSOs, and comparing these to their optical absolute $B$-band
magnitudes, it is clear that many of the X-ray detected QSOs lie close
to the limits of the X-ray data and hence the X-ray undetected QSOs may
represent the X-ray-faint wing of the distribution, rather than an
intrinsically different population.  This underlines the effectiveness
of deep X-ray surveys for identifying QSOs.

We find that none of our 17 new QSOs has detectable radio
emission brighter than 100$\mu$Jy at 1.4\,GHz (Simpson et al.\ 2006),
so they are all radio-quiet.

Finally, we have also compared the image profiles of the $K$-band light in the
QSOs to nearby stars.  We identify stars based on their $V\! JK$ colours in
Fig.~2 and select a subsample with the same range in $K$-band
magnitudes and lying within 120$''$ of the QSOs.  We stacked the images
of 17 QSOs and comparison stars to produce composite profiles and we plot
these in Fig.~5.  This shows that the QSOs are more spatially extended in
the $K$-band than nearby stars of similar magnitudes, with FWHM
measured from Moffat fits of $0.80\pm 0.03''$ for the QSOs and $0.67\pm
0.03''$ for the stars.  Subtracting a scaled star from the composite
QSO image suggest the extended component of the $K$-band light comprises
25--40\% of the light of the QSO, equivalent to $K_{tot}\sim 20.3$, and
appears to have a spatial extent of $\sim 2$--5\,kpc.  The crude sizes
and brightnesses of this component are similar to that measured from AO
imaging of high-redshift QSO hosts (e.g.\ Falomo et al.\ 2008) and the
fraction of the total light emitted in the restframe $R$-band is also
similar to that seen in local QSOs (Maddox \& Hewett 2006).  Thus this
extended component thus has many of the features expected for the QSO
host galaxy (Maddox \& Hewett 2006).

\subsection{Foreground Galaxies}

We show in Fig.~4 the colour distribution for the large sample of
foreground galaxies identified in our survey.  There are 302 galaxies
in our AAOmega sample with Q=1 spectra, their redshifts span
$z=0.03$--1.62 and a mean of $z=0.43\pm 0.22$.  The spectral mix of
this population includes sources whose spectra are best-fit by broad-line AGN
(53), narrow-line AGN (13), narrow emission-line galaxy (23),
absorption-line galaxy (75) and spiral galaxy (138) templates.  The
$(B-K)$ colour distribution for these galaxies extends to very red
colours, $(B-K)\gs 7$, and show an upper envelope consistent with the
expected colours of a non-evolving $L^*$ E/S0 galaxy (Fig.~4).

Eleven of these foreground galaxies, at redshifts of $z=0.20$--1.38,
are matched to the {\it XMM-Newton} X-ray source catalogue (Ueda et
al.\ 2008) within 5$''$. Ten of these 11 have spectra which are best
fit by broad-line (QSO) spectral templates, although their absolute
restframe $B$-band luminosities are fainter than $M_B\sim -23$.

\subsection{Improving the sample selection}

Our identification of 17 new QSOs from spectroscopy of over 400
sources indicates a modest rate of return.  This is mitigated
somewhat by the fact that we removed from our input catalogue those
QSOs already known from the existing spectroscopy of X-ray sources in
this field.  At best these would have doubled our success rate.
Nevertheless, it is clear that even with the grasp of the AAOmega
spectrograph a substantially larger future KX QSO survey would benefit
from reducing the contamination from foreground compact galaxies.

We have therefore investigated two routes we could have potentially
improved the yield of QSOs in our spectroscopic survey: 1) reducing the
limit on object compactness and 2) including a cut based on $(B-V)$
colour.  To illustrate this we show the distribution of $(B-V)$ colours
and $R$-band FWHM for the new QSOs from our AAOmega sample, the X-ray
selected QSOs in this field and the contaminating population of
foreground galaxies in Fig.~6.

A reduction in contamination could be achieved by adopting a $(B-V)\leq
1.2$ cut, which reduces contamination by $\sim 30$\% but at the cost of
potentially removing red QSOs and those at higher-redshifts.  Similarly
using $B$-band FWHM (cut at $\leq 1.1''$) can reduce contamination by
$\sim 40$\%.  Combining these two selections would reduce contamination
by $\sim 60$\% overall and result in a final QSO yield of 10\%.

Adopting a more rigorous limit on FWHM, say $\leq 1.0''$, would
significantly reduce the contamination, by $\sim 80$\%, at the cost of losing a
small number of QSOs with resolved hosts ($\sim 10$\%).  For certain experiments,
especially those targeting high-redshift QSOs, such an approach might be
appropriate as the host galaxies of such systems are unlikely to be
individually detected and in this mode KX-selection would yield very
high completeness.

We caution that our sample is far from ideal for identifying
an optimal selection as the selection criteria will be tuned for
selecting QSOs similar to those that we have found: at $z\ls 3.5$ and
relatively blue.  A better training set would be provided by a more
complete spectroscopic sample, including near-infrared spectroscopy
with FMOS, to overcome the bias towards optically bright QSOs in the
current sample.
We also note that further detailed studies of the contaminating
population may be a productive route to identify criteria to remove
them from the KX QSO sample.

%
%
\begin{figure}
\centerline{\psfig{file=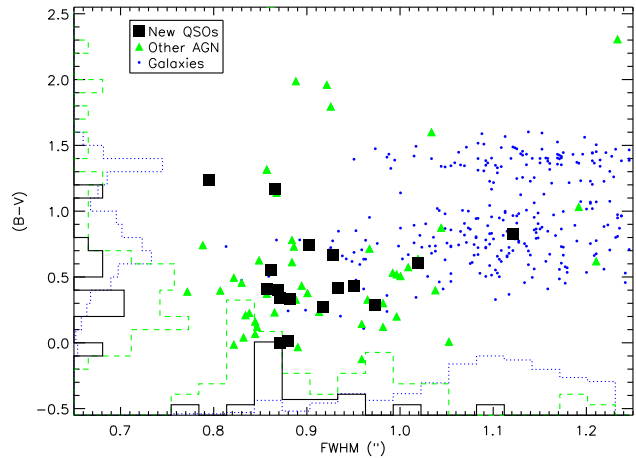,angle=90,width=3.5in}}
\caption{$(B-V)$ versus FWHM in the $R$-band for spectroscopically
identified QSOs in the UKIDSS UDS (both the X-ray and KX-selected) and
the contaminating population of foreground galaxies.  We plot
histograms showing the distribution in $(B-V)$ colour and FWHM for each
of the three samples: new KX QSOs (solid), existing AGN (dotted) and galaxies
(dashed).  These distributions illustrate the challenge of isolating
the QSO populations from the compact, faint galaxies.  We conclude that
complete samples of QSOs cannot be efficiently
constructed from simple selection based on image compactness or colour,
although more stringent limits on image compactness would yield
significant reductions in contamination, at relatively modest cost in
terms of incompleteness.
}
\end{figure}

\section{Conclusions}

We present the results from a pilot survey for QSOs in the UKIDSS UDS
field.  We employ the KX selection method of Warren et al.\ (2000) to
select a sample of faint candidate QSOs from the very deep optical and
near-infrared imaging available for this field.  We have then followed
up these sources with the 400-fibre AAOmega spectrograph in service
time on the AAT to identify the high-redshift QSOs in the $\sim
700$  KX-selected compact sources in our catalogue.
We identify 17 QSOs with M$_B\leq -23$ in this pilot survey.
Supplemented by existing spectroscopy of X-ray selected AGN in this
field (Akiyama et al.\ 2008) this yields a combined QSO sample with a
surface density of $\sim 85$--150\,degree$^{-2}$ at
$K_{tot}\leq 20$ (where the range
reflects the uncertainty in the spectroscopic incompleteness in our
sample).  This sample is roughly $4\times$ deeper than previous
KX-selected surveys over comparable areas (Jurek et al.\ 2008) and
$40\times$ larger than the previous surveys at similar $K$-band
depth (Croom et al.\ 2001).

We have used the good image quality of the UKIDSS UDS  data to
create composite $K$-band images of our QSO sample and a matched
sample of stars.  The light profiles of the two composites show
that the QSOs are more spatially extended.  The extended component
comprises some 25--40\% of the total $K$-band light of the QSO and has
a spatial extent of $\sim 2$--5\,kpc.  This most likely represents
the host galaxy of the QSO.

We analyse our sample to attempt to refine the KX selection technique
to reduce the considerable contamination by low redshift, $z\ls 0.6$,
compact galaxies.  We find that colour or morphological selections
focusing on bluer passbands (e.g.\ an additional $(B-V)$ cut or limits
on image FWHM in the $B$-band) would reduce the contamination in our 
parent sample.  However, these approaches would bias the resulting
sample against redder or higher-redshift QSOs and so may not be
desirable for certain applications where completeness is
important.  Alternatively, if the goal is to maximise
the number of high-redshift
QSOs detected, then the
KX selection using a more conservative FWHM limit, $\leq 1.0''$
in the case of the $0.8''$-FWHM UDS imaging,  can significantly
reduce the foreground contamination while at the same time yielding
high completeness ($\sim 90$\%).

From our sample we cull a subsample of $\sim 21$ QSOs at $z>2.5$ and $V<23$
which are suitable for an absorption line
study (see Adelberger et al.\ 2005) employing deep blue spectroscopy
with VIMOS on VLT.  In addition, the full sample of QSOs (including
those at lower redshifts) provide a map of the distribution of rapidly
growing black holes within our survey volume, enabling us to relate
these to the growth of mass-selected samples of galaxies and the
surrounding absorption-line systems.  These luminous AGN are expected
to have measurable feedback effects on the gas -- an issue of great
interest to current models of galaxy formation (in which such forms of
feedback are an essential part, Bower et al.\ 2006).

\section*{acknowledgments} 

We thank the members of the UDS Working Group, in particular Chris
Simpson, and Natasha Maddox, Paul Hewett and Dave Alexander for help
and useful discussions.  We thank the Referee for their clear
and insightful report, which helped improve the accuracy and
presentation of this paper.
IRS acknowledges support from the Royal
Society. AMS acknowledges support from STFC.  This work is based in
part on data obtained as part of the UKIRT Infrared Deep Sky Survey.
We acknowledge the contributions of the staff of UKIRT to the
implementation UKIDSS survey and the Cambridge Astronomical Survey Unit
and the Wide Field Astronomy Unit in Edinburgh for processing the
UKIDSS data.  The United Kingdom Infrared Telescope is operated by the
Joint Astronomy Centre on behalf of the Science and Technology
Facilities Council of the U.K.  This paper uses data obtained through
the Service Programme at the Anglo-Australian Telescope.  This work was
based in part on observations made with the Anglo-Australian
Telescope. We warmly thank all the present and former staff of the
Anglo-Australian Observatory for their work in building and operating
the 2dF and AAOmega facilities.


\begin{thebibliography}{}
\bibitem{} Adelberger, K., Shapley, A.E., Steidel, C.C., Pettini, M., Erb, D.K., Reddy, N.A., 2005, ApJ, 629, 636
\bibitem{} Akiyama, M., et al., 2008, in prep.
\bibitem{} Almaini, O., et al., 2008, in prep.
\bibitem{} Blanton, M.R., Roweis, S., 2007, AJ, 133, 734
\bibitem{} Bower, R.G., Benson, A.J., Malbon, R., Helly, J.C., Frenk, C.S., Baugh, C.M., Cole, S., Lacey, C.G., 2006, MNRAS, 370, 645
\bibitem{} Cannon, R., Drinkwater, M., Edge, A., Eisenstein, D., Nichol, R., Outram, P., Pimbblet, K., de Propris, R., Roseboom, I., 2006, MNRAS, 372, 425
\bibitem{} Cirasuolo, M., McLure, R.J., Dunlop, J.S., Almaini, O.,
  Foucaud, S., Smail, I., Sekiguchi, K., et al., 2007, MNRAS, 380, 585
\bibitem{} Colless, M.M., Dalton, G., Maddox, S., Sutherland, W.,
  Norberg, P.,  Cole, S., Bland-Hawthorn, J., Bridges, T.,
   et al., 2001, MNRAS, 328, 1039
\bibitem{} Croom, S.M., Warren, S.J., Glazebrook, K., 2001, MNRAS, 328,
  150
\bibitem{} Croom, S.M., et al., 2001, MNRAS, 322, L29
\bibitem{} van Dokkum, P., 2001, PASP, 113, 1420
\bibitem{} Falomo, R., Treves, A., Kotilainen, J., Scarpa, R., Uslenghi,
  M., 2008, ApJ, 673, 694
\bibitem{} Foucaud, S., Almaini, O., Smail, I., Conselice, C.J., Lane,
  K. P.; Edge, A.C., Simpson, C., Dunlop, J.S., et al., 2007, MNRAS, 376, L20
\bibitem{} Furusawa, H., Kosugi, G., Akiyama, M., Takata, T., Sekiguchi, K., Tanaka, I., Iwata,
I., Kajisawa,  M., et al., 2008, ApJS, 176, 1
\bibitem{} Glikman, E., Gregg, M.D., Lacy, M., Helfand, D.J., Becker, R.H., White, R.L., 2004, ApJ, 607, 60
\bibitem{} Glikman, E., Helfand, D.J., White, R.L., Becker, R.H., Gregg, M.D., Lacy, M., 2007, ApJ, 667, 673
\bibitem{} Jurek, R.J., Drinkwater, M.J., Francis, P.J., Pimbblet, K.A., 2008, MNRAS, 383, 673
\bibitem{} Lane, K., Almaini, O., Foucaud, S., Simpson, C., Smail, I.,
  McLure, R .J., Conselice, C.J., Cirasuolo, M., et al., 2007, MNRAS,  379, L25
\bibitem{} Lawrence, A., Warren, S.J., Almaini, O., Edge, A.C., 
Hambly, N.C., Jameson, R.F., Lucas, P., et al., 2007, MNRAS, 379, 1599
\bibitem{} Maddox, N., Hewett, P.C., 2006, MNRAS, 367, 717 
\bibitem{} Maddox, N., Hewett, P.C., Warren, S.J., Croom, S.M., 2008,
  MNRAS, 386, 1605 
\bibitem{} Morris, S., Januzzi, B., 2006, 367, 1261
\bibitem{} Prescott, M.K.M., Impey, C.D., Cool, R.D., Scoville, N.Z.,
  2006, ApJ, 644, 100
\bibitem{}Saunders, W., Bridges, T., 
Gillingham, P., Haynes, R.,  Smith, G.A., Whittard, J.D., 
Churilov, V., Lankshear, A., et al., 2004, SPIE, 5492, 389
\bibitem{} Sharp, R., et al., 2002, 337, 1153
\bibitem{} Sharp, R., Saunders, W., Smith, G., Churilov, V., Correll,
  D., Dawson, J., Farrel, T., Frost, G., et al., 2006, SPIE, 6269, 14

\bibitem{} Simpson, C., Martinez-Sansigre, A., Rawlings, S., Ivison, R.,
  Akiyama, M., Sekiguchi, K., Takata, T., Ueda, Y., Watson, M., 2006,
  MNRAS, 372, 741
\bibitem{} Smith, G.A., Saunders, W., Bridges, T., Churilov, V.,
  Lankshear, A.,; Dawson, J., Correll, D.,  et al., 2004, SPIE, 5492, 410
\bibitem{} Ueda, Y., et al., 2008, in prep
\bibitem{} Warren, S., Hewett, P.C., Foltz, C.B., 2000, MNRAS, 312, 827
\bibitem{} Warren, S., Hambly, N.C., Dye, S., Almaini, O., Cross,
  N.J.G., Edge, A.C., Foucaud, S., Hewett, P.C.,  et al., 2007, MNRAS, 375, 213

\end{thebibliography}
\end{document}